# Raman spectra of crystalline secondary amides

Boris A. Kolesov[a,b]

[a] *Institute of Inorganic Chemistry SB RAS, Novosibirsk, Russia;*
[b] *Novosibirsk State University, Novosibirsk, Russia; b.kolesov@g.nsu.ru*

## Abstract

The study of single-crystal Raman spectra of a series of crystalline secondary amides (acetanilide, methacetin, phenacetine, orthorhombic and monoclinic polymorphs of paracetamol) as well as simple amides formanilide and benzanilide in the temperature range 5-300 K was carried out. The series of compounds with the same molecular fragment - acetamide group – can serve as a model system to study the interrelation between the latter and the properties of the intermolecular "peptide-type" NH···O=C hydrogen bonds. For all the "acetamide family" of compounds, similar changes in the Raman spectra were observed on cooling the samples: an appearance of new Amide-I$^-$ and Amide-I$^+$ bands that are red and blue shifted respectively from the conventional Amide-I band by around of 5-10 $cm^{-1}$. An appropriated changes in the same temperature range were observed for the N-H out-of-plane bending (Amide-V) and N-H stretching vibrations of the N-H···O=C hydrogen bond. All the spectral changes on cooling the samples can be supposed to result from delocalization of the Amide-I and N-H modes and appearance of dynamical (Davydov's) splitting at low temperature.

**Key words**: molecular crystals, Raman spectra, secondary amides, hydrogen bonds, dynamical splitting

## Introduction

At the present time much research is dedicated to crystals of small organic molecules which are used as model systems. These crystals are also characterized by a network of hydrogen bonds, and any intramolecular motions and conformational changes in them may be strongly correlated with the cooperative motions of the entire dynamic network. The structures with acetamide group are especially interesting in this respect, since they have in the same molecular fragment a methyl-group, as well as the C=O and the N-H groups forming intermolecular hydrogen bonds, similar to those in peptides, the so called "peptide-type" hydrogen bonds. The acetamide group can be attached to different additional fragments, this giving rise to the "acetamide family" of compounds, including, among other members, acetanilide ($C_8H_9NO$), paracetamol ($C_8H_9NO_2$), methacetin ($C_9H_{11}NO_2$), phenacetine ($C_{10}H_{13}NO_2$) (Figure 1). In addition to the possibility to vary the environment of the acetamide-groups by modifying molecular structures, one can change juxtaposition of the same molecules in the crystal structure either in a continuous way (variations of temperature and pressure within the range of stability of the same phase), or more radically (by preparing different polymorphs). In the abovementioned series, polymorphs were reported for paracetamol.

In our previous publication [1] we have demonstrated (using the two polymorphs of paracetamol as examples), that one more technique, namely variable-temperature polarized Raman spectroscopy of single crystals in several well-defined crystallographic orientations, can be extremely informative, when studying the changes of molecular backbone conformations in relation to the changes in the crystal structure, in general, and in intermolecular hydrogen bonds, in particular. The state of the acetamide group at different temperatures (i.e. its conformation and the bonds parameters) are best of all characterized by the following Raman modes: stretching C=O vibrations at ~ 1650 $cm^{-1}$ (Amide-I), composite vibrational band (including the bending (in-plane) N-H and stretching C-N vibrations) with two components at ~1550 $cm^{-1}$ (Amide-II) and ~1250 $cm^{-1}$ (Amide-III). Another characteristic and informative band is the N-H bending out-of-plain vibration at 740-800 $cm^{-1}$ ($\gamma_{N-H}$ or Amide-V). The last but not least, the parameters of the N-H···O hydrogen bond itself in a series of related compounds can be correlated with the wavenumber of the stretching N-H vibration.

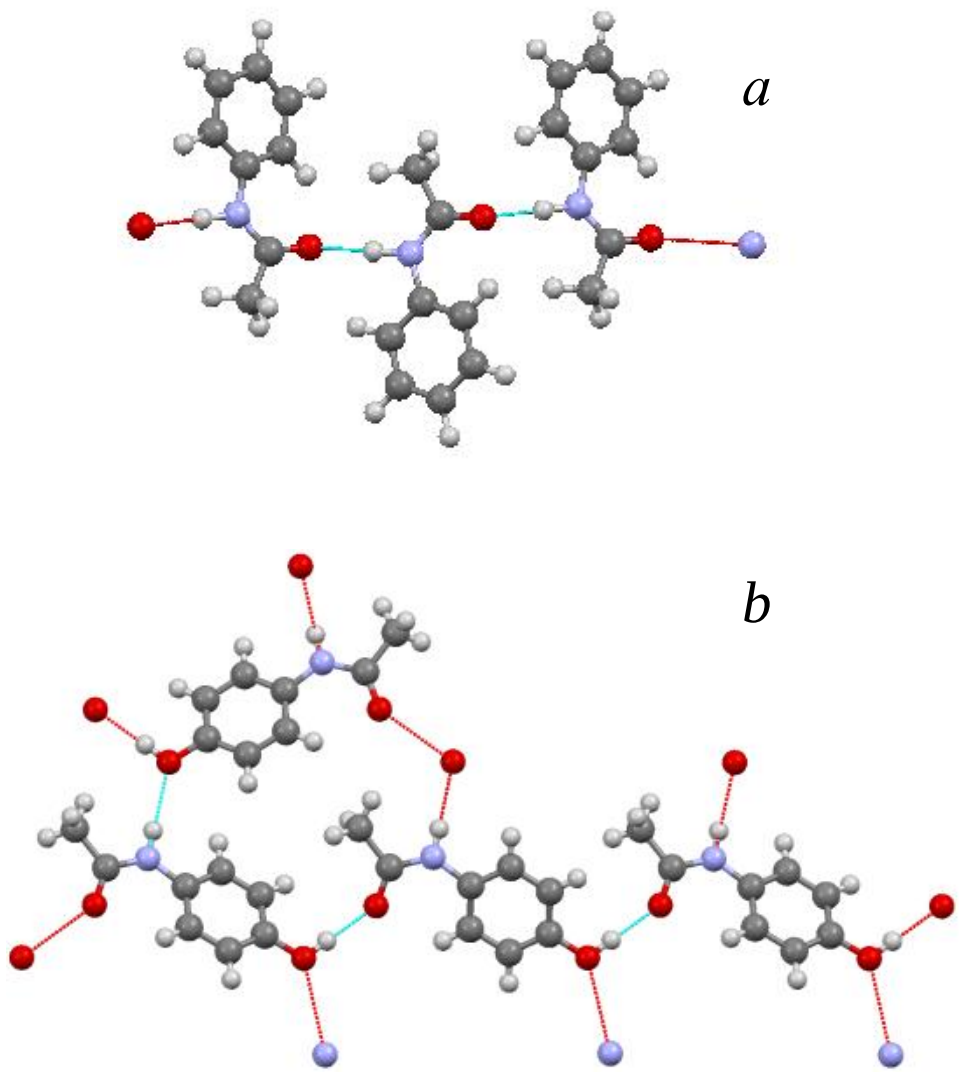

Fig. 1. The chains of hydrogen bonds in the structure of acetanilide (a) and chains and cycles in the structure of rhombic paracetamol (b).

One of the main result obtained in Ref. [1] for the monoclinic and the orthorhombic polymorphs of paracetamol was that the relative integral intensities of the two neighbouring modes at about 1660 $cm^{-1}$ changed oppositely in a correlated way on temperature variations: the intensity of the higher wavenumber mode decreased, whereas that of the lower wavenumber mode increased upon cooling, so that the total integral intensity of the two modes remained constant. We have assigned the two bands to C=O stretchings, and have interpreted the observed effect as a manifestation of the rotation of the methyl-group at about 60° around the C-C bond (*i.e.* of the change of the conformation from staggered to eclipsed with respect to C=O bond on cooling). To test this conclusion we carried out the Raman study of the other compounds with acetamide group in the present work.

The increase in the intensity of the low-wavenumber satellite of Amide-I on cooling was reported earlier for the crystals of acetanilide [2,3]. However, this effect was interpreted in terms of the generation of the self-trapped states (Davydov's solitons [4]) in the N-H···O=C bonded chains, after what several dozens of publications appeared, mainly the theoretical ones, discussing the self-trapping in the crystals of acetanilide. Therefore a study of an extended series of crystalline compounds with acetamide group was necessary.

*The aim of the present study* was to extend research to other members of the "acetamide family", namely, to acetanilide, methacetine and phenacetine, and to measure single-crystal polarized Raman spectra for them in the temperature range 5 K – 300 K. We aimed to test, if the same effect can be observed for all the structures, in order to make more general conclusions, on how the changes in molecular structure and crystal packing are related to acetamide group vibrations and N-H···O=C hydrogen bonding. The main attention was paid to the spectral changes in the "range of Amide I vibrations", i.e. 1630-1670 $cm^{-1}$, in which the pronounced effect of the redistribution of integral intensities between the bands was observed for the polymorphs of paracetamol.

## Experimental

Raman spectra were collected using a LabRAM HR Evolution, Horiba spectrometer with a CCD detector Symphony from Jobin Yvon. The 488 nm line of an Ar+ laser (35LAP431 from Melles Griot Company, USA) was used for the spectral excitation with a diameter of the laser spot on the sample surface of 2 μm. The laser power at the sample was typically 1 mW. The spectra at all temperatures were obtained in the backscattering geometry using a Raman microscope and helium cryostat of closed cycle (ARS Company, United States). The crystals were wrapped in indium foil for a better thermal contact, keeping an open field at the upper surface accessible for light, and fixed on a cold finger of the cryostat. All measurements were performed with a spectral resolution of 1.5 $cm^{-1}$.

## Results

In the temperature range 5 K – 300 K, Raman spectra of the crystals exhibit several features identical for all compounds of the acetamide family. Figure 2,a shows the spectra of acetanilide in the range of stretching vibrations of the hydrogen bonded C=O group. In really the spectrum at room temperature consists of dominating Amide I (1666 $cm^{-1}$) band and two satellites of weak intensity: Amide $I^-$ (1654 $cm^{-1}$) and Amide $I^+$ (1668 $cm^{-1}$). (Due to thermal broadening of Raman lines the bands at 1666 and 1668 $cm^{-1}$ are overlapped at elevated temperature and can be evaluated by deconvolution of the spectra into Lorentz components). At low temperature the modes show an inverse relation of their intensities (Fig. 2,a,c): strong Amide $I^-$ and Amide $I^+$ bands and very weak Amide I. The total intensity remains as a constant (Fig. 2,c). (Only two modes Amide $I^-$ and Amide $I^+$ were observed in IR-spectra of acetanilide at low temperature in Ref. [2,3]. A reason of disagreement between the spectral data at low temperature relates to an imperfection of the crystals used in this work).

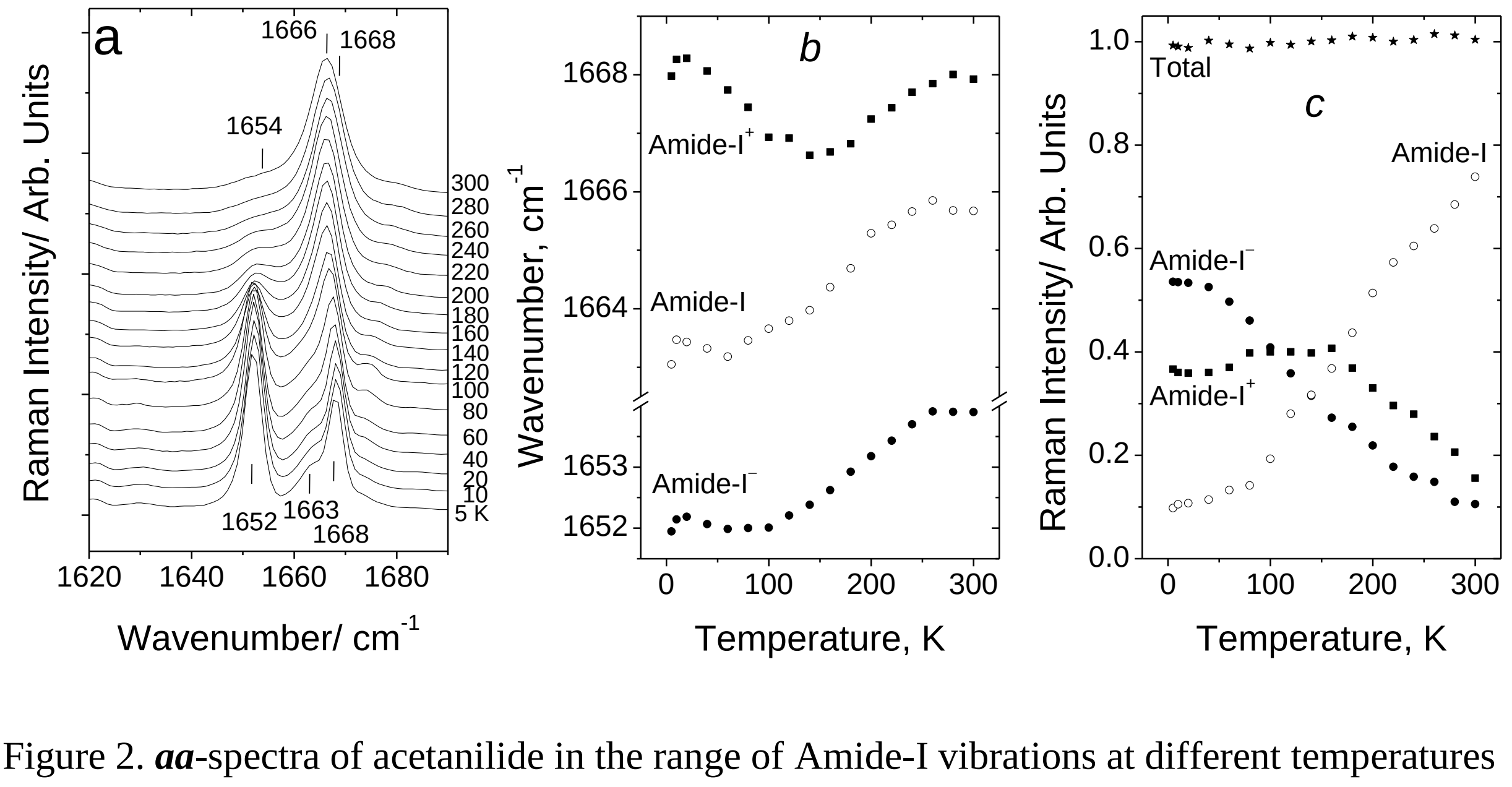

Figure 2. ***aa***-spectra of acetanilide in the range of Amide-I vibrations at different temperatures (a); Peak position (b) and integral intensity (c) of Amide-I (open circles) , Amide-$I^-$ (full circles), Amide-$I^+$ (full squares) and total intensity (asterisks) as function of temperature. Here and further the definitions ***aa***, ***bb*** and ***cc*** imply the directions of the polarization vector of the incident (first symbol) and the scattered (last symbol) light with respect to crystal axes.

Figure 3 shows a changing with temperature of the Amide-V mode (778 $cm^{-1}$) in acetanilide, typical for all the other compounds. The Amide-V mode, very weak at ambient temperature, increases in wavenumber and intensity at cooling down and this changing is synchronized in temperature with that observed for Amide-I, Amide-$I^-$ and Amide-$I^+$ (Figure 2).

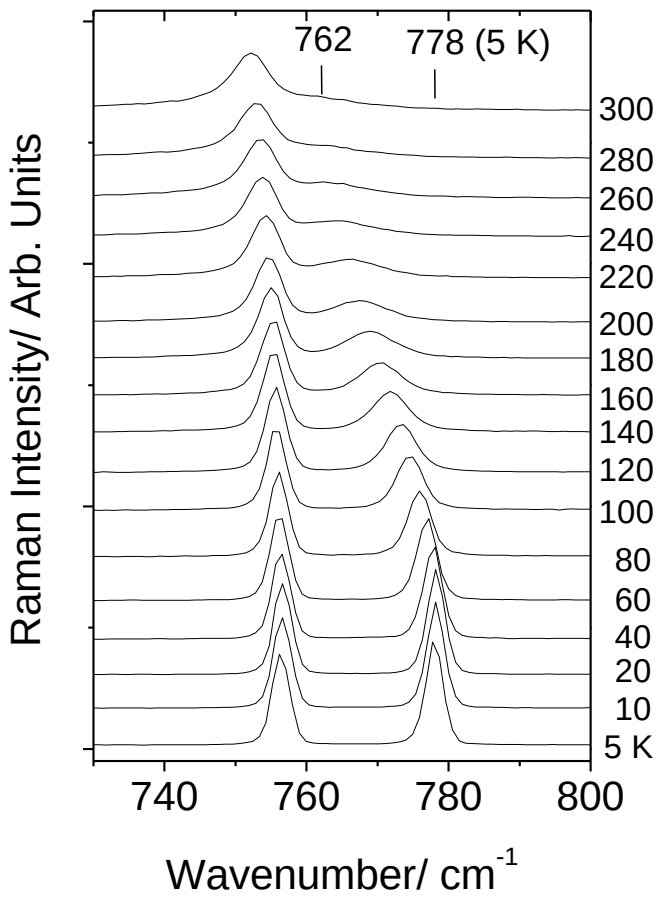

Fig. 3. ***aa***-spectra of acetanilide in the range of Amide-V vibrations ($\gamma_{N-H}$).

The most complicated variation with temperature occurs in the spectra of acetanilide, methacetin and phenacetine in the range of N-H stretching vibrations of N-H···O=C hydrogen bond. At room temperature two broad bands are observed at ~ 3264 and 3296 $cm^{-1}$ for acetanilide, Fig. 4,a. Upon cooling another two bands appear progressively resulting in four main bands, which are clearly resolved at 5 K (Fig. 4,a). In the case of paracetamol, rhombic and monoclinic, N-H vibration appears in the spectra as a single mode at around of 3330 $cm^{-1}$ at all temperatures (Fig. 4,b)

## Discussion

As it was already mentioned, the appearance of additional mode in the range of Amide-I in the spectra of acetanilide crystals was interpreted in the previous reports as an arising of self-trapped states [2,3]. However, the presence of two satellites, Amide $I^-$ and Amide $I^+$, instead of one in all the compounds with acetamide group excludes the suggestion on self-trapping. In fact, the modes related to the vibrations of C=O fragment as well as to N-H stretching and N-H out-of-plane bending show the correlated temperature behavior. In addition, the Amide $I^-$, Amide $I^+$ and Amide-V bands appear in the spectra of rhombic and monoclinic paracetamol in the same way as it occurs in the other compounds of acetamide family in spite of that the paracetamol is characterized by O-H···O=C hydrogen bonding.

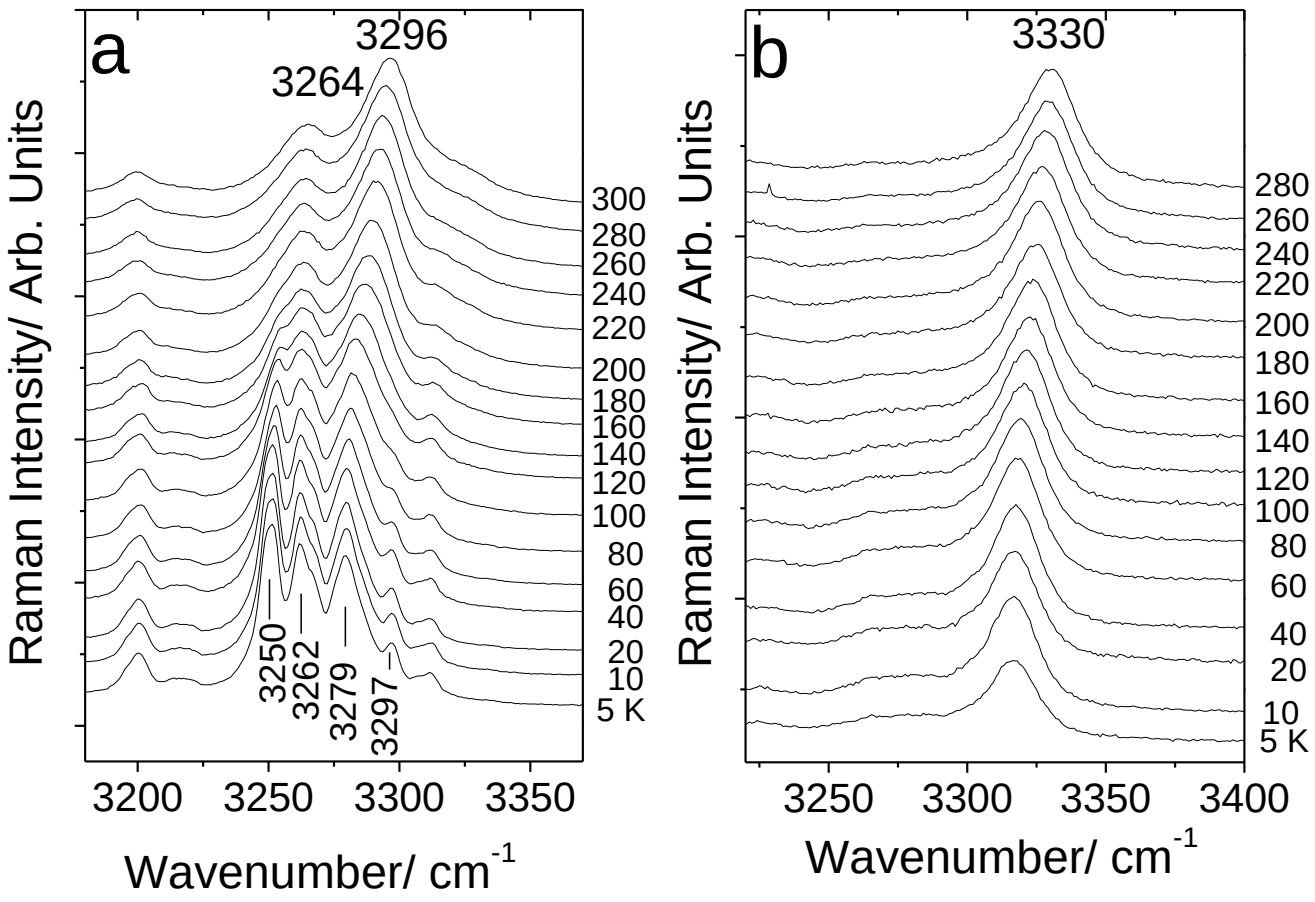

Fig. 4. ***bb***-spectra of acetanilide (a) and rhombic paracetamol (b) in the range of N-H stretchings.

On the other hand, the studies of the crystal structure at low temperature [5,6] show no any structural phase transitions except negligible changes of the intramolecular parameters, i.e. bond lengths and angles, which are common for all molecular crystals. It means that the observed spectral features in the 5-300 K range (Fig. 2-4) do not result from structural phase transitions which could be fixed in structural studies.

The presence in low-temperature spectrum of two main modes, Amide $I^-$ and Amide $I^+$, can be formally interpreted as existence of two different in energy states of N-H···O=C (O-H···O=C for paracetamol) hydrogen bonds. There are a few hypothetical reasons for appearance in the spectra of these states. The first one is a rotation of $CH_3$ group at about 60° around the C-C bond from staggered to eclipsed conformation with respect to C=O bond. Another imaginary reason is an arising of additional hydrogen bonding between C-H of aromatic cycle and oxygen atom of carbonyl O=C group. Finally, one can imagine an existing of two different positions of proton on N-H···O=C hydrogen bonding and so on. However, in all the cases the only states of minimal energy should be realized in the system at low temperature. In addition, the mentioned hypothetical processes should progress in the crystals as stepwise phase transitions.

Basing on these comments one have to suggest an arising of specific dynamical interaction in the chains of N-H···O=C (O-H···O=C) hydrogen bonds at low temperature as a main reason of the observed effects.

The O=C vibrations of the same chain are able to interact between them giving rise the symmetric (in phase) and asymmetric (in out-of-phase) modes. The interaction depends strongly on the hydrogen bond strength, which plays a role of connecting link between the oscillators. At room temperature, however, a thermal population of different vibrations weakens the link. At this an exited vibration of any O=C bond is characterized by accidental vibrational phase, the vibration localizes within the single molecule and Amide-I band preferentially is observed in the spectra. When the temperature goes down the distance between the molecules shortens due to the freezing of low-frequency vibrations, especially of donor-acceptor translational modes. At this the phases of the O=C vibration of the neighbor molecules become correlate between them. In other words, the vibrations delocalize and become look like phonons. In the perfect crystals only two modes, symmetric and asymmetric, realizes at low temperature. In the crystals where the chains are distorted in some extent at any temperature the localized vibrations (third mode) can be conserved what is observed in the crystals used in this work.

The splitting of O=C vibrations takes place for the molecules of the same chain only in spite of that the total number of O=C groups in the unit cell can be more than 2. (For example, there are six O=C groups in the unit cell of acetanilide). It makes different the dynamical splitting in hydrogen bonded chains with the common dynamical (Davydov's) splitting in crystals. In addition the temperature-dependent process of mode delocalization is also a distinct feature of the hydrogen bonded chains.

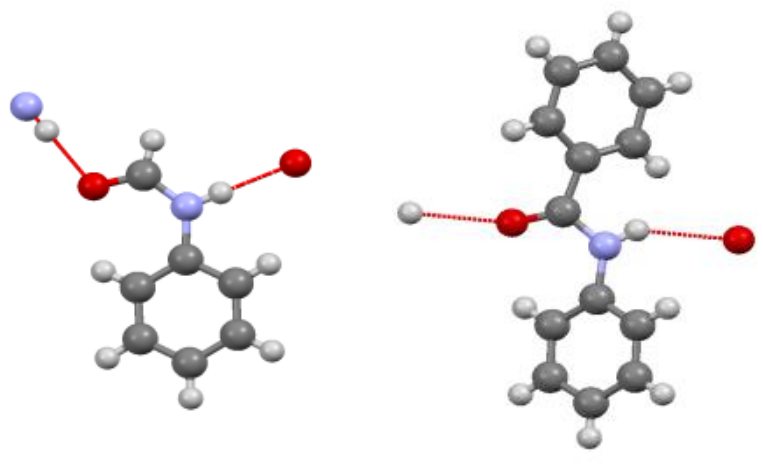

*a* *b*

Fig. 5. N–H···O=C chains in the structure of formanilide (a) and benzanilide (b).

In order to substantiate the suggestion above, the spectra of formanilide ($C_7H_7NO$) and benzanilide ($C_{13}H_{11}NO$) crystals (Fig. 5) were taken at different temperatures. N-H···O=C hydrogen bonding is significantly stronger in formanilide and weaker – in benzanilide as compared with the bonding in the crystals of acetamide family. It was found that the only Amide-I$^-$ and Amide-I$^+$ modes are presented in the spectra of formanilide, and the only Amide-I mode – in the spectra of benzanilide at all temperatures. The main characteristics of hydrogen bonding and spectral data of all the compounds are listed in Table 1. Fig. 6,a shows the splitting Δω (a difference between Amide-I$^+$ and Amide-I$^-$ mode wavenumber) at 5 K as function of $d_{N\cdots O}$ distance. The dependence was plotted on the base of data listed in Table I. One can see that Δω tents to zero at $d_{N\cdots O}$ around of 2.98 Å. The N···O distance in acetanilide and phenacetine is less than 2.98 Å by just 0.03-0.04 Å that is comparable with the vibrational amplitude of thermally excited lattice vibrations. Thus the phenomenon of vibrational localization at ambient temperature and delocalization – at low temperature results from a thermal population of intermolecular vibrations and takes place only for the sufficiently narrow interval of the donor-acceptor distances. The compounds of acetamide family just fall into this interval.

**Table 1**

**Hydrogen bond length, frequency of bending Amide-V and stretching of Amide-I$^-$, Amide-I$^+$ vibrations at 5 K and difference Δω between two latter modes**

| Compound | Type of H-bonding | D···A, Å | Amide-V, cm$^{-1}$ | Amide-I$^-$, cm$^{-1}$ | Amide-I$^+$, cm$^{-1}$ | Δω, cm$^{-1}$ |
|---|---|---|---|---|---|---|
| Formanilide | N-H···O=C | 2.84 [7] | 802 | 1647 | 1699 | 52 |
| Methacetin | N-H···O=C | 2.91 [8] | 802 | 1647 | 1674 | 27 |
| Acetanilide | N-H···O=C | 2.94 [9] | 778 | 1652 | 1668 | 16 |
| Phenacetine | N-H···O=C | 2.95 [10] | 768 | 1648 | 1664 | 16 |
| Benzanilide | N-H···O=C | 3.15 [11] | 701 | 1652 | | 0 |
| Paracetamol monoclinic | O-H···O=C | 2.65 [12] | 751 | 1637 | 1648 | 11 |
| Paracetamol rhombic | O-H···O=C | 2.72 [6] | 738 | 1638 | 1650 | 12 |

Fig. 6,b shows the Δω in acetanilide as function of temperature. The dependence is fitted well by thermal excitation of phonon with wavenumber of ~85 cm$^{-1}$ (solid curve). This frequency is typical one for intermolecular vibrations of hydrogen bonds, that is vibrations, which effect on temperature change of $d_{N\cdots O}$ in much extent.

If oscillators (i.e. O=C) of force constant $K$ and frequency ω are linked between them by weak springs of force constant $k$, than $\Delta\omega/\omega \sim k/K$, where Δω is a difference between symmetric and asymmetric modes of the oscillators. For ω=1660 cm$^{-1}$ and Δω=16 cm$^{-1}$ (acetanilide, Table 1), $k/K \sim 0.01$. It is very realistic relation between force constants of O=C and N···O bonds.

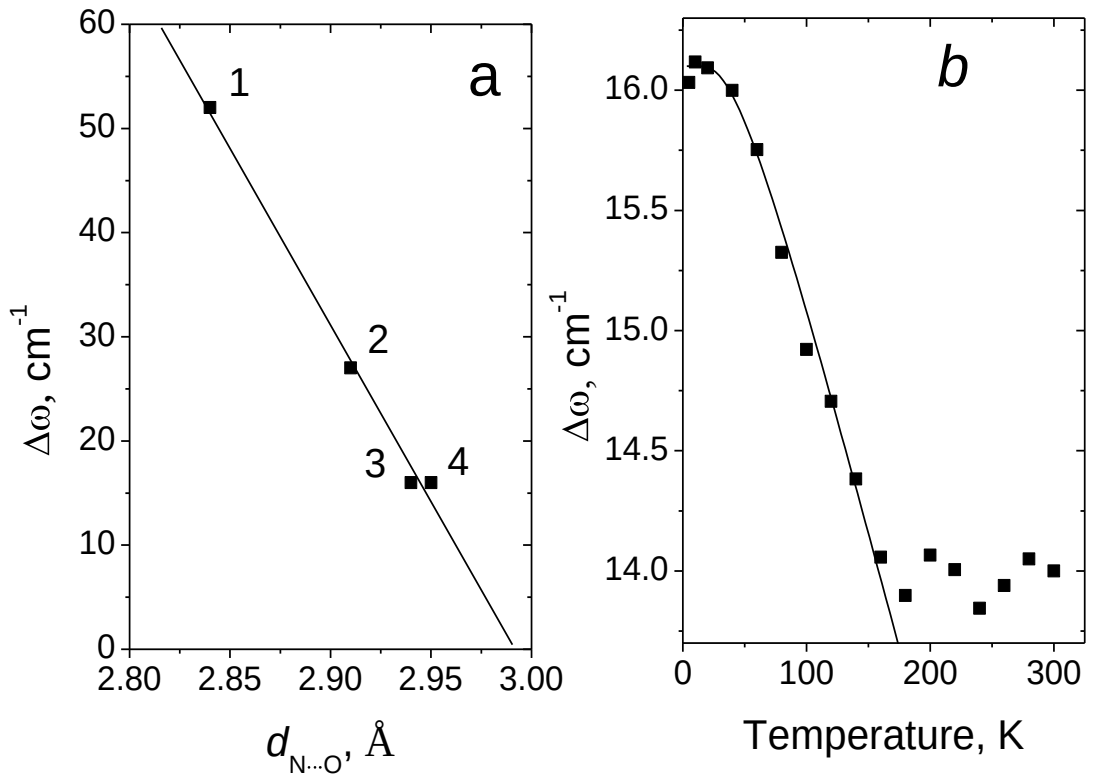

Fig. 6. (a): The splitting Δω = ω (Amide-I$^+$) – ω(Amide-I$^-$) as function of N···O length, 1 – formanilide, 2 – methacetin, 3 – acetanilide, 4 – phenacetine; (b) The same as function of temperature in acetanilide; solid curve is associated with a freezing (exciting) of phonon of 85 cm$^{-1}$ wavenumber.

It is well known that N-H···O=C (O-H···O=C) hydrogen bonding (and, thus, Amide-I frequency) depends strongly on a chemical nature of the substituent R in secondary amides. In the compounds of acetamide family R=$CH_3$ and methyl group fixes the hydrogen bond characteristics within the narrow interval. It makes possible the observation of the same spectral features for all the compounds. As for the supposed rotation of $CH_3$ group, which could provoke a conformational transition, the spectra of formanilide evidences a non-participation of $CH_3$ in the splitting of Amide-I.

The spectroscopic manifestation of the N-H stretching vibrations as function of temperature is rather complicated (as compared with the occurrence of Amide-$I^-$ and Amide-$I^+$) and requires additional comments. First of all, the number of the observed bands is surprising, namely two main bands are in the spectra at ambient temperature, and four bands – at low temperatures. Granting that each band is of direct relevance to a separate hydrogen bond, several hydrogen bonds of different strength should be expected to co-exist in the crystal structures, what is, however, not the case. This is why an alternative interpretation of the effects observed for NH-vibrations has to be suggested.

There is only single N-H···O=C hydrogen bond in the crystal structure with the localized N-H stretches in the range of 3200-3300 $cm^{-1}$, but the overtone of the Amide-I (~1650 $cm^{-1}$) also falls into this wavenumber range. Since N-H and C=O fragments are involved into the same N-H···O=C hydrogen bond, the $\omega_{N-H}$ and $2\omega_{O=C}$ vibrations are coupled strongly providing a Fermi-resonance pair, which is represented in the spectra at room temperature by two broad bands at 3264 and 3296 $cm^{-1}$ (Fig. 4,a). At temperature decreasing the localized N-H stretching becomes delocalized (i.e. the dynamical splitting becomes available) in the same way as the Amide-I does. At this each of two components of dynamically splitted N-H exhibits a Fermi-resonance with an overtone of Amide-$I^-$ or Amide-$I^+$, as a result of which two pairs of Fermi-resonanced modes appear in the spectra (Fig. 4,a). A change in the N-H···O=C hydrogen bond strength in the studied series reveals to the corresponding change in the C=O stretching wavenumber, so that the described Fermi-resonance is observed for all the compounds with the exception of paracetamol.

In monoclinic and orthorhombic paracetamol, N-H···O hydrogen bonding is formed between N-H and hydroxyl –OH (not O=C) and no Fermi resonance is observed in the Raman spectra (Fig. 4b). One more hydrogen bond, O-H···O=C, produces in the spectra of orthorhombic paracetamol several broad and overlapped bands[1], which can be assigned to different combined tones. The O-H stretches of O-H···O=C bond in the spectra of monoclinic paracetamol are too weak in the intensity (hardly measurable).

Finally it needs to mention one more inherent characteristic of a chain of hydrogen bonds. The Amide-V bending ($\gamma_{N-H}$) does not demonstrate a dynamical splitting at all temperatures (Fig. 3) in contrast to O=C stretching. It occurs because of that the stretching O=C modulates the length of N-H···O=C hydrogen bond while the N-H bending does not. It makes interaction between Amide-V vibrations of neighbor molecules negligible, not observed experimentally. A different behavior of stretching and bending motions in a chain of hydrogen bonds was reported for $H_2O$ chains (one-dimensional ice) in cavities of bikitaite mineral [13].

## Conclusion

The present variable-temperature Raman spectroscopic study of a series of related crystalline compounds with acetamide group has revealed some interesting features, which can be interpreted assuming that the chain of hydrogen bonds provides an interaction of those vibrations, which modulate the hydrogen bond length. The interaction starts up a delocalization of the vibrational modes at low temperature. The suggestion has permitted to justify the phenomenon of O=C stretch splitting, which was assigned to a self-trapping for a long time. This phenomenon can be of interest when considering the mechanisms of phonon propagation in molecular crystals and the physical properties of molecular materials including biopolymers, for example, of peptides and proteins. The diffraction data for the latter are usually collected at low temperatures (100-200 K), and it is important to be able to predict the changes in the hydrogen bonds and the lattice dynamic which can result from cooling as compared to those at ambient conditions.

**Acknowledgements**

We thank Prof. E. Boldyreva from the Novosibirsk State University for providing the crystals of acetanilide, methacetin and phenacetine. We also thank Prof. A. Tikhonov from the Novosibirsk Institute of Organic Chemistry for providing the crystals of formanilide and benzanilide.

## Figure captures

Fig. 1. The chains of hydrogen bonds in the structure of acetanilide (a) and chains and cycles in the structure of rhombic paracetamol (b).

Fig. 2. ***aa***-spectra of acetanilide in the range of Amide-I vibrations at different temperatures (a); Peak position (b) and integral intensity (c) of Amide-I (open circles) , Amide-$I^-$ (full circles), Amide-$I^+$ (full squares) and total intensity (asterisks) as function of temperature. Here and further the definitions ***aa***, ***bb*** and ***cc*** imply the directions of the polarization vector of the incident (first symbol) and the scattered (last symbol) light with respect to crystal axes.

Fig. 3. ***aa***-spectra of acetanilide in the range of Amide-V vibrations ($\gamma_{N-H}$).

Fig. 4. ***bb***-spectra of acetanilide (a) and rhombic paracetamol (b) in the range of N-H stretchings.

Fig. 5. N–H···O=C chains in the structure of formanilide (a) and benzanilide (b).

Fig. 6. (a): The splitting $\Delta\omega = \omega$ (Amide-$I^+$) – $\omega$(Amide-$I^-$) as function of N···O length, 1 – formanilide, 2 – methacetin, 3 – acetanilide, 4 – phenacetine; (b) The same as function of temperature in acetanilide; solid curve is associated with a freezing (exciting) of phonon of 85 $cm^{-1}$ wavenumber.